\documentclass[
    ,final            
  ]
  {aipproc}

\layoutstyle{6x9}

\begin{document}

\title{Electromagnetic Structure of the Pion}



\author{Clayton S. Mello}{
  address={Laborat\'orio de F\'{\i}sica Te\'orica e Computacional (LFTC), \\Universidade Cruzeiro do Sul, 
01506-000, S\~ao Paulo, Brazil}
}

\author{Jos\'e P. Cruz Filho}{
}

\author{Edson O. da Silva}{
}

\author{Bruno~El-Bennich}{
}

\author{J. P. B. C. de Melo}{
}

\author{Victo S. Filho}{
}

\begin{abstract}
In this work, we analyze the electromagnetic structure of the pion. We calculate its electromagnetic radius and electromagnetic form factor in low and intermediate momentum range. Such observables are determined by means of a theoretical model that takes into account the constituent quark and antiquark of the pion within the formalism of light-front field theory. In particular, we consider a nonsymmetrical vertex in this model, with which we  calculate the electromagnetic form factor of the pion in an optimized way, so that we obtain a value closer to the experimental charge radius of the pion. The theoretical calculations are also compared with the most recent experimental data involving the pion electromagnetic form factor and the results show very good agreement. 
\end{abstract}

\maketitle

\section{Introduction}

In experimental investigations of the pion, a great deal of experimental data concerning its electromagnetic structure, as its electromagnetic form factor $F_{\pi}(q^2)$, has been reported~\cite{Data3,Data2,Data2a,Huber,Tadevosyan,Brauel,Horn}.
In the theory of light mesons, relevant problems considered in the literature concern, for instance, the calculation of the mass and decay constant of   the pion in a relativistic potential model of independent quarks~\cite{Jena}, the study of masses and electroweak properties of light mesons in a relativistic quark model~\cite{Faustov} and the study of a relativistic treatment of pion wave functions in the annihilation $\bar{p}p \rightarrow \pi  \pi^+$~\cite{Bruno}, beside a lot of other works in which are considered calculations involving light mesons, as in Refs.~\cite{LC09,Roberts,Maris,Tsirova,Leitner11}. 

One of the theoretical ways for describing so many experimental data is to adopt the light-front field theory formalism~\cite{Brodsky}. A model inspired by QCD was proposed in Ref.~\cite{Miller}. In the present work, we consider a particular description based on a nonsymmetrical vertex model, as reported in Ref.~\cite{Pacheco99}. In it, the light-front formalism is applied to the electromagnetic form factor, calculated with the $+$ components of the currents of the quark-antiquark bound states of the pion. 
Methods based on the light-front formalism have been successful in the description of the electromagnetic properties of the hadronic wave functions~\cite{Miller,Pacheco99,Choi,Dziembowski87,Cardarelli96,Simula2002,Jaus99,Hwang2003,Huang2004}. 

In the following, the light-front model with the nonsymmetrical vertex (NSV model) for the pion is briefly described, followed by the numerical results for the pion electromagnetic form factor and electromagnetic radius. 
\section{The Model}

The pion electromagnetic form factor in light-front field theory can be performed in the covariant form as: 
\begin{equation}
(p+p^{\prime})^{\mu} 
F_{\pi}(q^2)\ = \ <\pi(p^{\prime})|J^{\mu}|\pi(p)>, 
\label{ffactor}
\end{equation}
in which 
$q=p^\prime - p$ and the matrix elements of the electromagnetic current $J^{\mu} = e(p^{\mu}+p^{\prime\mu})F_{\pi}(q^2)$ are given by
\begin{eqnarray}
J^{\mu} = - 2  \imath  e N_c \frac{m^2}{f^2_\pi} \int \frac{d^4k}{(2\pi)^4} Tr \Bigl[ S(k) \gamma^5 S(k-p^{\prime})\gamma^\mu S(k-p) \gamma^5 \Bigr]\Gamma(k,p^{\prime})\Gamma(k,p), 
\label{current}
\end{eqnarray} 
where 
$S(p)=(\rlap\slash p-m+\imath \epsilon)^{-1}$ is the quark propagator and $N_c=3$ is the number of colors. We adopt in our calculations the Breit frame, by considering initial momenta $p^{\mu}=(p_0,-q/2,0,0)$, final momenta $p^{\prime {\mu}}=(p_0,q/2,0,0)$ and $q^{\mu}=(0,q,0,0)$. 
As described in Refs.~\cite{LC09,Pacheco99}, the electromagnetic form factor of the pion receives only a valence contribution to the plus component of the electromagnetic current. In the case of the NSV model, $\Gamma(k,p)$ is the regulator vertex function known as nonsymmetrical vertex, which can be written as~\cite{LC09,Pacheco99}:
\begin{equation}
\Gamma^{NSY}(k,p)=\frac{N}{(p-k)^2-m^2_R+\imath\epsilon}
\label{nosymm} \  .
\end{equation} 
The form factor $F_{\pi}^{(NSY)}(q^2)$ in the NSV model, using the $+$ component of the electromagnetic current, can be expressed with the light-front wave function, as shown in Refs.~\cite{LC09,Pacheco99}, according with:
\begin{eqnarray}
F_{\pi}^{(NSY)} = \frac{m^2}{p^+ f^2_\pi} N_c \int \frac{d^{2} k_{\perp} d x}{2(2 \pi)^3 x } {\cal{N}}(x,p^+)
 \Psi^{*(NSY)}_f(x,k_{\perp}) 
\Psi^{(NSY)}_i(x,k_{\perp}) \theta(x) \theta(1-x), 
\label{form}
\end{eqnarray}
in which ${\cal{N}}(x,p^+)=-4 \frac{f_1}{x p^+}(x p^+ - p^+)^2 + 4 f_1(x p^+-2 p^+) + \ x p^+ q^2$, 
$f_1=k_{\perp}^2+m^2$ and $x=k^{+}/p^{+}$ is the fraction of the carried momentum by the quark.
The light-front wave function with the nonsymmetric vertex can be written as: 
\begin{equation}
\Psi^{(NSY)}(x,k_{\perp})=\frac{N}{(1-x)^2 (m_{\pi}^2-{\cal M}_0^2) (m_{\pi}^2-{\cal M}_R^2)}. 
\label{wavefunction}
\end{equation}
Here, ${\cal M}_R^2={\cal M}^2(m^2,m^2_R)= \frac{k_{\perp}^2+m^2}{x}+\frac{(p-k)_{\perp}^2+m_R^2}{(1-x)}-p^2_{\perp}$ and 
${\cal M}^2_0={\cal M}^2(m^2,m^2)$ is the free mass operator. The normalization constant $N$ obeys the condition $F_{\pi}(0)=1$. 
\section{Numerical Results}
In the numerical calculations, we initially considered the parameters $m_{u}=m_{\bar{d}}=0.220$ GeV, $m_R = 0.946$ GeV and $ m_{\pi} = 0.140$ GeV. For the input data cited, we obtained for the pion charge radius the value $\langle r_{\pi^+} \rangle \cong 0.689$ fm, with an error about 2\% smaller than the experimental value ($\langle r^{exp}_{\pi^+} \rangle = (0.672 \pm 0.008)$ fm)~\cite{PDG}. In order to improve the description of the experimental data, we explored the variation of the regulator mass $m_R$ and studied its influence on the form factor for low and intermediate momentum range. 
\vskip -.25cm
\begin{figure}[ht]
\includegraphics[angle=-90, scale=.4]{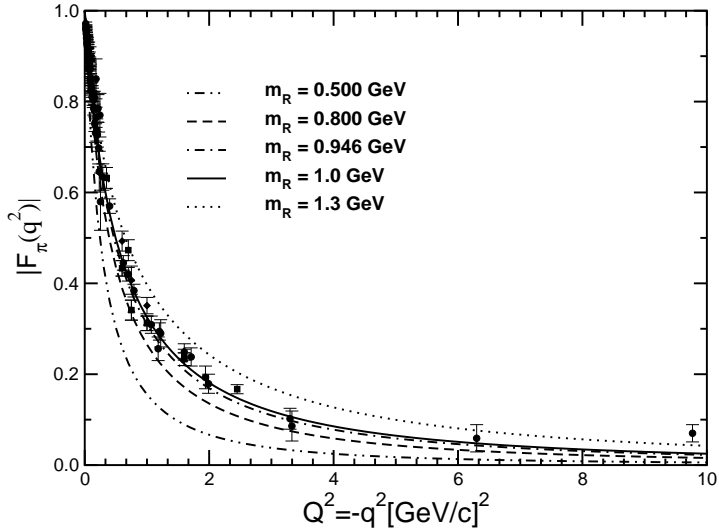}
\caption{Plot of the electromagnetic form factor versus $Q^2=-q^2 ([GeV/c]^2)$, calculated in NSV model. The different curves correspond to the electromagnetic form factor for some value of regulator mass, that is, $m_R$=$0.5, 0.8, 0.946, 1.0$ and $1.3$ GeV. The light constituent quark masses are $m_u$=$m_{\bar{d}}=0.220$ (GeV). The numerical results are compared with the experimental data, as described in Refs.~\cite{Data3,Brauel} (full triangle up), Ref.~\cite{Huber} (full square), Ref.~\cite{Tadevosyan} (full circle), and Ref.~\cite{Horn} (full diamond).}
\label{lcp1}
\end{figure}

We show in figure \ref{lcp1} the numerical results obtained in the NSV model for the electromagnetic form factor of the pion, up to 10~$[GeV/c]^2$. By analyzing the curves in figure \ref{lcp1}, we conclude that $m_R = 1.0$ GeV provides the best value to describe with more precision the experimental data~\cite{Data3,Huber,Tadevosyan,Brauel,Horn}. Values of the regulator mass smaller than $m_R =  0.8$ GeV and above $m_R = 1.3$ GeV do not show good agreement with the experimental data for a wide range of momenta. In order to confirm the best value of $m_R$, one can calculate the electromagnetic radius. In fact, we have calculated it for all of the values considered and, for $m_R = 1.0$ GeV, we obtain $\langle r_{\pi^+} \rangle \cong 0.673$ fm, with a 0.2\% deviation of the experimental value. In order to check our model, we also calculated the pion decay constant for $m_R = 1.0$ GeV, obtaining $f_{\pi} = $ 93.1 MeV, very close to the experimental value $f^{exp}_{\pi} = $ 92.2 MeV. 
\vskip -.2cm
\section{Conclusions}
\vskip -.2cm
We conclude that the best value of the regulator mass for the nonsymmetrical vertex model in the light-front formalism is $m_R = $ 1.0 GeV. With that value, it is possible to describe with the best precision the experimental data for the electromagnetic form factor of pion. We also studied the dependence of the model on the regulator mass. The numerical results show that the model significantly breaks down for $m_R < $ 0.8 GeV and also fails for higher values, above 1.3 GeV. 

\begin{theacknowledgments}

\hspace{20cm} . 

We thank \ the brazilian agencies \ CAPES ({\it Coordenadoria de Aperfei\c{c}omento de Pessoal de N\'\i vel Superior}), CNPq ({\it Conselho Nacional de Desenvolvimento
Cient\'\i fico e Tecnol\'ogico})  and  FAPESP ({\it Funda\c{c}\~ao de Amparo \`a Pesquisa do Estado de S\~ao Paulo}), for financial support.
\end{theacknowledgments}

\bibliographystyle{aipproc}   




\IfFileExists{\jobname.bbl}{}
 {\typeout{}
  \typeout{******************************************}
  \typeout{** Please run "bibtex \jobname" to optain}
  \typeout{** the bibliography and then re-run LaTeX}
  \typeout{** twice to fix the references!}
  \typeout{******************************************}
  \typeout{}
 }

\end{document}